# A novel method for identification of local conformational changes in proteins

## Naoto Morikawa

**Genocript, Zama-shi, 228-0001, Japan**

**ABSTRACT**

**Motivation:** Proteins are known to undergo conformational changes in the course of their functions. The changes in conformation are often attributable to a small fraction of residues within the protein. Therefore identification of these variable regions is important for an understanding of protein function.

**Results:** We propose a novel method for identification of local conformational changes in proteins. In our method, backbone conformations are encoded into a sequence of letters from a 16-letter alphabet (called $D^2$ codes) to perform structural comparison. Since we do not use clustering analysis to encode local structures, the $D^2$ codes not only provides a intuitively understandable description of protein structures, but also covers wide varieties of distortions. This paper shows that the $D^2$ codes are better correlated with changes in the dihedral angles than a structural alphabet and a secondary structure description. In the case of the N37S mutant of HIV-1 protease, local conformational changes were captured by the $D^2$ coding method more accurately than other methods. The $D^2$ coding also provided a reliable representation of the difference between NMR models of an HIV-1 protease mutant.

**Availability:** Program ProteinEncoder and detailed data are freely available at http://www.genocript.com.

**Contact:** nmorika@genocript.com

## 1 INTRODUCTION

Proteins are known to undergo conformational changes in the course of their functions. The changes in conformation may range from small local fluctuations to large domain motions. But they are often attributable to a small fraction of residues within the protein. Therefore an efficient and accurate algorithm for the identification of conformational changes is crucial for protein function analysis.

Over the past decade, numerous methods have been developed for flexible structural alignment. However current flexible alignment methods, such as RAPIDO (Mosca *et al.*, 2008), FATCAT (Ye and Godzik, 2003), FlexProt (Shatsky *et al.*, 2002), and DynDom (Hayward *et al.*, 1997) often ignore subtle local differences between conformations.

To quantify the structural differences between local backbone conformations, a great variety of methods have been proposed. In those studies, the backbone conformation is often described using various geometric descriptors, such as, backbone dihedral angles ($\varphi$, $\psi$), dihedral angles of the C$\alpha$ trace, or distances derived from the positions of C$\alpha$ atoms (Korn and Rose, 1994; Kuznetsov and Rackovsky, 2003; Miao *et al.*, 2008; Flocco and Mowbray, 1995; Francisco *et al.*, 2007; Kelley *et al.*, 1997; Schneider, 2000; Rackovsky and Scheraga, 1984) Using these methods, one can pinpoint differences between two conformations of the same protein. But their accuracy are limited by coordinate errors associated with crystal structures, distortions due to crystal packing, and others. As for ($\varphi$, $\psi$), significant fluctuations of neighbouring dihedral angles do not necessarily mean a change in the C$\alpha$ trace because of the anticorrelation between them (McCammon *et al.*, 1977).

Another approach for quantification of structural differences is based on secondary structure assignments, where the variable regions of a protein are identified as the regions of variable secondary structure (Chou and Fasman, 1974). However secondary structures in proteins usually have a number of distortions such as bends, twists, and end fraying. In other words, proteins can change their conformation without changing their secondary structure.

To capture local backbone conformations more accurately, "structural alphabet" has been proposed in various studies, where descriptions by the traditional three-state secondary structure are replaced with descriptions by a library of local structure prototypes (de Brevern *et al.*, 2000; de Brevern *et al.*, 2001; Kolodny *et al.*, 2002; Camproux *et al.*, 2004; Wang *et al.*, 2005; Yang and Tung, 2006; Pandini *et al.*, 2010). A structural alphabet is usually obtained by clustering protein backbone fragments based on their conformations, where local protein structures are described differently by different groups. According to Le *et al.*. (2009), 1D representations based on a structural alphabet perform significantly better than the 1D representation based on the secondary-structure for fold recognition purposes, but not enough to become a viable replacement to computationally intensive procedures such as structural alignment tools.

Finally, Chang *et al.*. (2006) proposed a local geometric measure, the "writhing number," which is obtained from the knot theory. By encoding the distribution of writhing numbers across all the structures in the PDB, protein geometries are represented in a 20-letter alphabet.

In the following we propose a simple encoding method which translates the conformation of the C$\alpha$ trace into a sequence of letters from a 16-letter alphabet (called $D^2$ codes), which is based on the discrete differential geometry of tetrahedron sequences. In our method, five-residue fragments are first approximated by a tetrahedron sequence. Then, the fragments are assigned one of the 16 letters, using the second derivative of the corresponding tetrahedron sequence. Since we do not use clustering analysis to encode local structures, this approach not only provides a intuitively understandable description of protein structures, but also covers wide varieties of distortions. Regarding structural alignment, one can use naive sequence alignment algorithms to compare protein structures. In *SHREC'10 Protein Model Classification* we achieved results comparable to more sophisticated methods, using the length of the longest common subsequence as the measure of structural similarity (Mavridis *et al*., 2010).

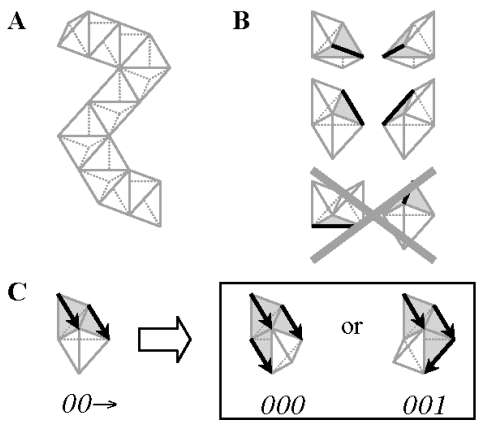

**Fig. 1.** Sequence of tetrahedra. (**A**) Example of tetrahedron sequence. (**B**) Six ways of connecting three tetrahedra. We use only four of them, excluding two U-turns (bottom). (**C**) Encoding of variation in a gradient vector along a tetrahedron sequence. The binary sequences (shown underneath) describe the variation in a gradient vector of the gray blocks.

## 2 METHODS

### 2.1 The $D^2$ coding of backbone conformations

In contrast to previous studies, we don't use the position of individual atoms nor secondary structure to identify local structural features. Instead, we approximate the backbone conformation by a folded tetrahedron sequence and consider the second derivative of the tetrahedron sequence.

*2.1.1 Folded tetrahedron sequence* To encode local protein structure, we use space filling tetrahedra. Each of the tetrahedra consists of four short edges and two long edges, where the ratio of their length is $\sqrt{(3)}/2$. Connecting tetrahedra which share a face one after another, we obtain a folded tetrahedron sequence (Figure 1A). Although there are six ways to connect three tetrahedra (Figure 1B), we forbid U-turns (Figure 1B, bottom). As a consequence, there are only two ways to add a new tetrahedron to the tail of a tetrahedron sequence. For instance, let's consider a three-tetrahedron sequence in Figure 1C (left), where the tail tetrahedron is colored white. Shown in the right-hand panel of Figure 1C are the two tetrahedra (colored white) that can be added to the tail.

To encode the conformation of a tetrahedron sequence, we consider the "gradient" of a tetrahedron, which is the direction of the edge of the tetrahedron that is not shared with the adjacent tetrahedra (Bold lines in Figure 1B). Note that a "current" tetrahedron assumes one of two gradients: the same gradient as the "predecessor" or the other. We use bold arrows to indicate the gradient of a tetrahedron, as in Figure 1C.

Now we can describe variation in a gradient (i.e., the "second derivative") along a tetrahedron sequence by a {0, 1}-valued sequence. That is, change the value if the gradient is changed. Suppose that the previous tet-rahedron is assigned a value of 0 (Figure 1C, left). Then, assign a value of 0 to the current tetrahedron if the tetrahedron has the same gradient as the predecessor (left in the right panel of Figure 1C). Otherwise, assign a value of 1 to the current tetrahedron (right in the right panel). In the case of the three-tetrahedron sequence of Figure 1C, we obtain a binary sequence of 000 or 001, depending on the gradient of the tail tetrahedron.

*2.1.2 Encoding algorithm* In the first place, each of the five-residue fragments in a protein is approximated by a five-tetrahedron sequence. To absorb the irregularity inherent in actual protein structures, we allow translation and rotation of tetrahedra in the process as described below. Next, we compute the second derivative of the tetrahedron sequences to obtain {0, 1}-valued sequences of length five. In this study {0, 1}-valued sequences are denoted as a base-32 number:0, 1, ..., 9, A, B, ..., V. For example, 00010 is denoted by "2", 01001 is denoted by "9", 01010 is denoted by "A", and so on. Then, we assign the base-32 numbers to the center $C\alpha$ atom of the corresponding five-residue fragment. By arranging the base-32 numbers in the order they appear in the $C\alpha$ trace, we obtain a 1D representation of protein structures. We call the base-32 number sequence the "$D^2$ code" sequence of a protein.

As an illustration, let's consider the $C\alpha$ fragment $C(i-1)C(i)C(i+1)C(i+2)$ shown in Figure 2A. The gradient of the $C\alpha$ fragment at the i-th $C\alpha$ atom $C(i)$ is defined as the direction from the position of $C(i-1)$ to that of $C(i+1)$ (Rackovsky and Scheraga, 1978). In the following, we denote the vector from point A to point B by AB.

The initial tetrahedron T(i), OPQR in Figure 2b, is given as follows. T(i) is aligned with $C(i)$ in such a way that (1) the gradient of T(i) is parallel to the gradient of the $C\alpha$ trace at $C(i)$, and (2) the direction of the vector OS + OP and the vector $C(i)C(i-1) + C(i)C(i+1)$ coincide, where S = (O + RQ) is a vertex of an aligned tetrahedron. By definition T(i) is always assigned the gradient type shown on the left in the middle row of Figure 1C. For size, the length of the shorter edges is about one-fifth of the average distance between the consecutive $C\alpha$ atoms. (A small change in the length does not affect the outcome.)

Once the spatial orientation of T(i) is fixed, the position and spatial orientation of the next tetrahedron T(i+1) are determined uniquely. On the other hand, we have two choices for the gradient of T(i+1), which are shown in Figure 2C and 2D. Then the gradient of T(i+1) is determined based on the distance between T(i+2) and C(i+1). In the current case, we choose the gradient shown in Figure 2C (Figure 2E) because T(i+2) in Figure 2C is closer to $C(i+1)$ than T(i+2) in Figure 2D. Since the gradient of T(i+1) is different from that of T(i), a value of 1 is assigned to T(i+1).

Next, T(i+1) is moved along the $C\alpha$ trace to absorb structural irregularities of proteins: T(i+1) is translated to the position of $C(i+1)$ (Figure 2F) and rotated there (Figure 2G). To be more precise, T(i+1) is rotated around the cross-product $Grad(T(i+1)) \times Grad(C(i+1))$ until the direction of $Grad(T(i+1))$ coincides with that of $Dir(C(i+1))$ (i.e. "turn" without "twist", where $Grad(T(i+1))$ is the gradient of T(i+1) and $Grad(C(i+1))$ is the gradient of the $C\alpha$ trace at $C(i+1)$.

Once the spatial orientation of T(i+1) is fixed, the position and spatial orientation of the next tetrahedron T(i+2) are uniquely determined. But we have two choices for the gradient of T(i+2). The gradient of T(i+2) is determined based on the distance between T(i+2) and C(i+2) (Figure 2H). As the gradient T(i+2) is different from that of T(i+1), a value of 0 is assigned to T(i+2).

Then, we obtain the binary sequence of 010, which describes the variation in a gradient along the fragment $C(i)C(i+1)C(i+2)$. Encoding the fragment $C(i)C(i-1)C(i-2)$ starting from $C(i)$ in the same way, we obtain a {0, 1}-valued sequence of length five.

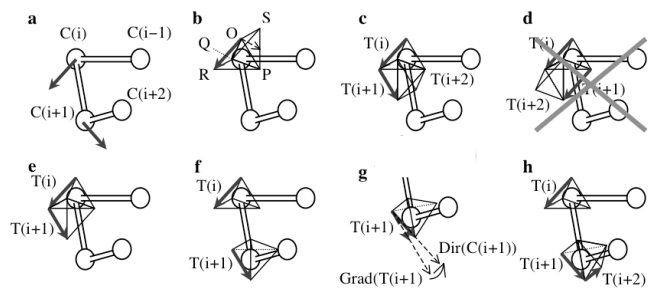

**Fig. 2.** The $D^2$ encoding algorithm. (**A**) The $C\alpha$ trace of a protein to be encoded. The arrows indicate the direction of a gradient vector of the $C\alpha$ trace. (**B**) Spatial orientation of the initial tetrahedron. (**C**), (**D**) Two permitted values of the gradient vector of $T(i+1)$. (**E**) The gradient vector of the first two tetrahedrons. (**F**) Translation of $T(i+1)$ to the position of $C(i+1)$. (**G**) Rotation of $T(i+1)$ in the position of $C(i+1)$. (**H**) The gradient vector of the three tetrahedrons.

### 2.1.3 Identification of variable regions

Using alternative experimental models of the same protein, conformationally variable residue positions are identified as the positions that assume more than one type of the $D^2$ codes, the PB blocks, or the DSSP assignments.

When more than two alternative experimental models are available, we define the "major" $D^2$ code for each residue position as follows. The "major" $D^2$ code of a residue position is the most frequently occurred $D^2$ code at the position among the models. The other $D^2$ codes occurred at the position are called "minor" $D^2$ codes. The major and minor PB blocks of a residue position are also defined similarly.

## 2.2 Performance evaluation protocol

We used an evaluation protocol described below to compare the performance of our method with three existing methods: the PB coding (de Brevern *et al.*., 2000) the DSSP coding (Kabsch and Sander, 1983), and a ($\varphi$, $\psi$) dihedral angle-based method (Kuznetsov, 2008). Assignments of the $D^2$ code and the PB block (PB code) were carried out with program ProteinEncoder and the PBE server (Tyagi *et al.*., 2006), respectively. DSSP state assignments (DSSP codes) were obtained from the PDB database of EMBL-EBI (http://www.ebi.ac.uk/msd/). We also used the DSSPcont server 8Carter *et al.*., 2003) to compute the DSSP assignment if not available from the database.

### 2.2.1 Three gold standards

We used backbone dihedral angles ($\varphi$, $\psi$) to define "gold standards" of conformationally variable residue positions as Kuznetsov (2008) did. To identify significant changes in the dihedral angles, Kuznetsov calculated standard deviations (31.8° for $\varphi$ and 34.8° for $\psi$) and used them as a cutoff value. In this study, we used 30.0° as a cutoff value for both $\varphi$ and $\psi$ for reasons of simplicity. Three kinds of residue positions are considered:

(1) The ANGL regions consist of the residue positions which satisfy either difference of $\varphi \geq 30°$ or difference of $\psi \geq 30°$.

(2) The ANGL_SUPP regions consist of the residue positions which satisfy either difference of $\varphi > 0°$ or difference of $\psi > 0°$.

(3) The ANGL_CORE regions consist of the residue positions which satisfy either (a) or (b), where (a) difference of $\varphi \geq 100°$ or difference of $\psi \geq 100°$, and (b) difference of $\varphi \geq 30°$ and difference of $\psi \geq 30°$.

The ANGL regions give the residue positions with variable dihedral angles. The ANGL_SUPP regions are the "support" of the variable region, i.e., all the variable positions should be contained in the region. And the ANGL_CORE regions are the "core" of the variable region.

### 2.2.2 Benchmark datasets

Three datasets are employed for the comparison study. The first is the set of 60 multiple-structure proteins identified by Kosloff and Kolodny (Kosloff and Kolodny, 2008). The dataset was chosen because the protein structures were aligned based solely on sequence information. The second is a set of 72 crystal structures of the N37S mutant of HIV-1 protease (PR), which is a homodimeric molecule, consisted of two identical 99-residue chains. (See supplement 1 for the PDBIDs of the 72 structures). HIV-1 PR is one of the major anti-HIV-1 drug targets (Wlodawer and Vondrasek, 1998) and a large collection of crystal structures of its variants are available in the PDB. The third is the 28 NMR models of an HIV-1 PR mutant [PDB:1bve]. Note that NMR structures are more prone to local distortions than X-ray structures because they correspond to proteins in solution. Therefore, they provide interesting cases for local structure encoding methods.

### 2.2.3 Performance measures

To measure the performance of each method, we calculated accuracy (ACC), sensitivity (SN), specificity (SP), and Matthews correlation coefficient (MCC) using the following formulas: ACC is $(TP + TN) / (TP + FP + TN + FN)$, SN is $TP / (TP + FN)$, SP is $TN / (TN + FP)$, and MCC is $(TP \cdot TN - FP \cdot FN) / \sqrt{(TP+FN)(TP+FP)(TN+FP)(TN+FN)}$, where TP, TN, FP, and FN are the numbers of true positives, true negatives, false positives, and false negatives, respectively.

## 3 RESULTS AND DISCUSSION

### 3.1 Conformationally variable regions

First the $D^2$ coding method was compared with the PB coding and the DSSP coding, using the 60 structure pairs identified by Kosloff and Kolodny and the there gold standards.

On average, 53.8% of all the residue positions in a protein undergoes changes in the dihedral angles (the ANGL_SUPP regions). Significant changes in the dihedral angles are observed in 17.8% of all the positions (the ANGL regions). Very large changes in the dihedral angles are observed in 10.2% of all the positions (the ANGL_CORE regions). On the other hand, 15.4% of all the residue positions assume multiple $D^2$ codes, 18.0% assume multiple DSSP codes, and 19.8% assume multiple PB codes.

The $D^2$ coding method is most specific with respect to the ANGL_SUPP regions. The PB coding method is most sensitive with respect to the ANGL_CORE regions. And the $D^2$ coding is most accurate with respect to the ANGL regions. Concerning MCC, the PB codes correlated better with the ANGL_CORE regions than others. The $D^2$ codes correlated better with the ANGL regions than others. But none of the three codes correlated well with the ANGL_SUPP regions. See supplement 2 for the ACC, SN, SP, and MCC computed against each of the three gold standards.

Recall that large changes in the dihedral angles do not necessarily mean an obvious change in the $C\alpha$ trace because of the anti-correlation between $\varphi$ and $\psi$. In particular, parts of the ANGL_CORE regions are not included in the $D^2$, PB, and DSSP variable regions. Therefore the values of MCC against the ANGL_CORE regions are not high. On the other hand, a residue position without any change in the dihedral angles may assume multiple D² or PB codes because they are computed from the conformation of a fragment centered on the residue. This is reflected in their low MCCs with respect to the ANGL_SUPP regions.

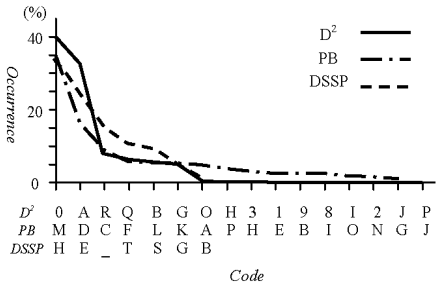

**Fig. 3.** The frequency of occurrence of *7* DSSP states, *16* D² codes, and *16* PB fragments among nine superfolds (1thbA, 256bA, 1aps, 1ubq, 2fox, 7timA, 1ilb, 2buk, and 2rhe).

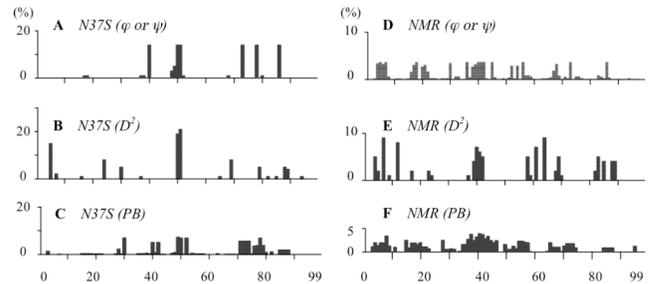

**Fig. 4.** The spatial distribution of the occurrence of (**A**) large deviation (>30°) in φ or ψ from the average, (**B**) minor D² code assignment, and (**C**) minor PB code assignment in the case of 72 crystal structures of the N37S mutant. The spatial distribution of the occurrence of (**D**) large deviation (>30°) in φ or ψ from the average, (**E**) minor D² code assignment, and (**F**) minor PB code assignment in the case of 28 NMR modes of a HIV-1 PR mutant.

Since the D² encoding is based on the discrete differential geometry of tetrahedron sequences, each of 16 D² codes has a intuitive meaning and they cover wide varieties of distortions. That is, "0"(=00000) is extended strand, "A"(=01010) is helix, "B"(=01011) is helix C-cap, "Q"(=11010) is helix N-cap, "R"(=11011) is turn, and the others correspond to varieties of distortions. In contrast, the letters in structural alphabets, such as the PB codes, often lack intuitive meaning and distortions are not well represented in the alphabets. That is because they are obtained by clustering fragments of protein backbone based on their conformations. In particular, both helices and extended strands correspond to a number of codes. Therefore one should compute a scoring matrix for substitutions to align structural alphabet sequences. Figure 3 shows the frequency distribution of the D², PB, and DSSP codes.

Consequently the D² variable region is usually much smaller than the PB variable region (15.4% < 17.8%(ANGL) < 19.8%). But it was shown that the D² codes correlated better with the ANGL regions than the PB codes (44.9% > 43.6%). As a result, one can perform D² code-sequence alignment without a substitution matrix. In *SHREC'10 Protein Model Classification* (Mavridis, 2010), we achieved results comparable to more sophisticated methods, using the length of the longest common subsequence (LCS) as the measure of structural similarity.

## 3.2 Difference between mutants

Next the D² coding method was compared with the PB coding and a (φ, ψ) dihedral angle-based method, using 72 crystal structures (142 chains) of the N37S mutant.

We first computed the major D² (and PB) codes for each residue position, which describe a typical local conformation of the 142 chains. Then we considered the spatial distribution of the minor D² (and PB) codes along the backbone (Figure 4).

On average, there are 2.5 minor D² codes, 7.1 minor PB codes, and 7.3 residues with large deviation (>30°) in φ or ψ per chain. About 40% of the minor D² codes are assigned at residue 50 or 51, and about 15% are assigned at residue 5. The deformation at residue 50 and 51 is explained by variation of the size of the ligands bound. The deformation at residue 4 is provably explained by crystal packing. As for the PB coding, not only around residue 50, but also residue 30, 40, and 71-80 often assume a minor PB code. On the other hand, the φ or ψ angle of residues 40, 50, 51, 73, 78, and 86 are almost always deviated from the average more than 30°.

As examples, let's consider the local structure of 1ajx chain A and chain B around residue 40 and residue 73 (Figure 5).

The φ angles of residue 40 of 1ajxA and 1ajxB are 133.7° and 101.5°, respectively. That is, the difference of their φ angles is 32.2°. The difference of their ψ angles is 15.0°. According to the alignment computed by the DALI server, the fragments from residue 25 to residue 55 of 1ajxA and 1ajxB are in very similar conformations (RMSD 0.5Angstrom). Therefore the change in dihedral angles does not represent a change in the Cα backbone.

The two chains have an identical D² code sequence ("00000"), but different PB code assignments ("EHIAD" and "EHJAC") around residue 40.

The φ angles of residue 73 of 1ajxA and 1ajxB are 177.6° and -170.2° respectively. Because of periodicity, the difference between them is 12.2°. According to the alignment computed by the DALI server, the fragments from residue 58 to residue 88 of 1ajxA and 1ajxB are in also very similar conformations (RMSD 0.3Angstrom). The fragments have an identical D² code sequence ("00000"), but different PB code assignments ("EHJAC" and "DFBLC") around residue 73.

In both cases, the D² code sequence captured the local conformation more accurately than the others. Regarding the (φ, ψ) dihedral angle-based method, it did not provide a reliable representation of protein local structures.

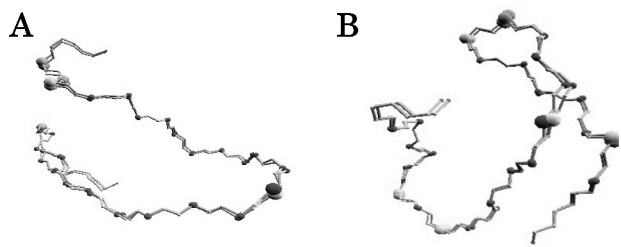

**Fig. 5.** Alignment of fragment pairs computed by the DALI server: (**A**) Res.25-55 of 1ajxA and 1ajxB, and (**B**) Res.58-88 of 1ajxA and 1ajxB. Large black spheres at the center of the fragments are residue 40 and 73 of 1ajxA. Large white spheres at the center are residue 40 and 73 of 1ajxB. The figures are prepared with program ProteinViewer.

### 3.3 Difference between NMR models

Lastly the $D^2$ coding method was compared with the PB coding and the $(\varphi, \psi)$ dihedral angle-based method used above, using 28 NMR models (56 chains) of a liganded HIV-1 protease.

On average, there are 8.6 minor $D^2$ codes, 19.2 minor PB codes, and 26.8 residues with large deviation (>30°) in $\varphi$ or $\psi$ per chain. In particular, more than one-fourth of all the residues undergo large deviation in the dihedral angles. Therefore it is difficult to characterize the difference between the NMR models in terms of the dihedral angles.

Concerning the $D^2$ coding, about 22% are assigned at the loop around residue 40, 9% are assigned at the loop around residue 7, and 9% are assigned at the loop around residue 64. The deformations are provably explained by collision with another molecule. Note that no minor $D^2$ code is assigned at the loop around residue 50, which is reasonable because all the NMR models have the same ligand. On the other hand, minority PB codes are assigned not only at the loops around residue 40 and around residue 7, but also at residue 51 and 52.

### 4 CONCLUSIONS

We proposed a novel method for identification of local conformational changes in proteins. In our method, protein backbone conformations are encoded into a sequence of letters from a 16-letter alphabet (called $D^2$ codes) to perform structural comparison. Since we do not use clustering analysis to encode local structures, the $D^2$ codes not only provides an intuitively understandable description of protein structures, but also covers wide varieties of distortions.

Three datasets are employed to compare the performance of our method with three existing coding methods: a structural alphabet (the PB blocks), a secondary structure description (the DSSP assignments), and a dihedral angle-based method. The first dataset is 60 multiple-structure proteins identified by Kosloff and Kolodny (2008). The second is 72 crystal structures of the N37S mutant of HIV-1 protease (PR). And the third is 28 NMR models of a HIV-1 PR mutant.

The results show that the $D^2$ codes are better correlated with changes in the dihedral angles than the PB blocks and the DSSP assignments although the $D^2$ coding method identifies much smaller variable regions than the other methods. In the case of the N37S mutant of HIV-1 PR, the local conformational changes were captured by the $D^2$ coding more accurately than the others. The $D^2$ coding method also provided a reliable representation of the difference between the dynamic NMR models of a HIV-1 PR mutant.

*Conflict of Interest*: none declared.